\documentclass{emulateapj}
\usepackage{apjfonts}
\usepackage{epsf}
\begin{document}
\slugcomment{}
\shortauthors{J. M. Miller et al.}
\shorttitle{A NICER Look at MAXI J1535$-$571}

\title{A {\it NICER} Spectrum of MAXI J1535$-$571: Near-Maximal Black Hole
  Spin and Potential Disk Warping}

\author{J.~M.~Miller\altaffilmark{1},
K.~Gendreau\altaffilmark{2},
R.~M.~Ludlam\altaffilmark{1},
A.~C.~Fabian\altaffilmark{3},
D.~Altamirano\altaffilmark{4}
Z.~Arzoumanian\altaffilmark{2},
P.~M.~Bult\altaffilmark{2},
E.~M.~Cackett\altaffilmark{5},
J.~Homan\altaffilmark{6,7},
E.~Kara\altaffilmark{8},
J.~Neilsen\altaffilmark{9},
R.~A.~Remillard\altaffilmark{10},
F.~Tombesi\altaffilmark{8,2,11,12}
}

\altaffiltext{1}{Department of Astronomy, University of Michigan, 1085
  South University Avenue, Ann Arbor, MI 48109-1107, USA,
  jonmm@umich.edu}
\altaffiltext{2}{NASA Goddard Space Flight Center, Code 662,
  Greenbelt, MD, 20771, USA}
\altaffiltext{3}{Institute of Astronomy, University of Cambridge, CB3 OHA, UK}
\altaffiltext{4}{Department of Physics \& Astronomy, University of
  Southampton, Southampton, Hampshire S017 1BJ, UK}
\altaffiltext{5}{Department of Physics \& Astronomy, Wayne State
  University, 666 W. Hancock Street, Detroit, MI, 48201, USA}
\altaffiltext{6}{Eureka Scientific, Inc., 2452 Delmer Street, Oakland
  CA 94602, USA}
\altaffiltext{7}{SRON, Netherlands Institute for Space Research,
  Sorbonnelaan 2, 3584 CA Utrecht, The Netherlands}
\altaffiltext{8}{Department of Astronomy, University of Maryland,
  College Park, MD 20742, USA}
\altaffiltext{9}{Department of Physics, Villanova University,
  Villanova, PA, 19085, USA}
\altaffiltext{10}{MIT Kavli Institute for Astrophysics and Space
  Research, 70 Vassar Street, Cambridge, MA 02139, USA}
\altaffiltext{11}{Department of Physics, University of Rome ``Tor
  Vergata'', Via della Ricerca Scientifica 1, I-00133 Rome, Italy}
\altaffiltext{12}{INAF Astronomical Observatory of Rome, Via Frascati
  33, 00078 Monteporzio Catone, Italy}
  
\begin{abstract}
We report on a {\it NICER} observation of the Galactic X-ray binary
and stellar-mass black hole candidate, MAXI J1535$-$571.  The source
was likely observed in an ``intermediate'' or ``very high'' state,
with important contributions from both an accretion disk and hard
X-ray corona.  The 2.3-10~keV spectrum shows clear hallmarks of
relativistic disk reflection.  Fits with a suitable model strongly
indicate a near-maximal spin parameter of $a = cJ/GM^{2} = 0.994(2)$
and a disk that extends close to the innermost stable circular orbit,
$r/r_{ISCO} = 1.08(8)$ ($1\sigma$ statistical errors).  In addition to
the relativistic spectrum from the innermost disk, a relatively {\it
  narrow} Fe~K emission line is also required.  The resolution of {\it
  NICER} reveals that the narrow line may be asymmetric, indicating a
specific range of emission radii.  Fits with a relativistic line model
suggest an inner radius of $r = 144^{+140}_{-60}~GM/c^{2}$ for the
putative second reflection geometry; full reflection models suggest
that radii a few times larger are possible.  The origin of the narrow
line is uncertain but a warp likely provides the most physically
plausible explanation.  We discuss our results in terms of the
potential for {\it NICER} to reveal new features of the inner and
intermediate accretion disk around black holes.
\end{abstract}

\section{Introduction}
MAXI J1535$-$571 was discovered as a new X-ray source by MAXI on 2017
September 02 (Negoro et al.\ 2017a; also see Nakahira et al.\ 2018).
The source displays many X-ray spectral and timing features consistent
with X-ray binaries that harbor a stellar-mass black hole, including
quasi-periodic oscillations (QPOs) and a relativistic Fe~K line
suggestive of a high black hole spin parameter ($a =
0.88^{+0.10}_{-0.20}$, where $a = cJ/GM^{2}$; Gendreau et al.\ 2017).
No Type-I X-ray bursts have been observed from MAXI J1535$-$571, nor
have any coherent pulsations or kHz quasi-periodic oscillations
(Negoro et al.\ 2017b, Gendreau et al.\ 2017); positive detections of
pulsations or kHz QPOs would signal a neutron star primary.  The radio
flux of the source also signals that it is likely a black hole
(Russell et al.\ 2017); neutron stars tend to have a lower radio/X-ray
ratio than stellar-mass black holes (e.g., Migliari \& Fender 2006).

The interstellar absorption along the line of sight to MAXI J1535-571
appears to be relatively high.  Values of $N_{H} = 3.6\pm
0.2\times 10^{22}~ {\rm cm}^{-2}$ have been reported (Kennea et
al.\ 2017).  Guver \& Ozel (2009) find that $N_{H} ({\rm cm}^{-2}) =
2.21\pm 0.09\times 10^{21}~ A_{V}$ (mag), this implies $A_{V} \simeq
16$.  Standard optical methods of measuring the mass function may be
very difficult for MAXI J1535$-$571.  

The most complete prior X-ray analysis of MAXI J1535$-$571 was made by
Xu et al.\ (2018), based on a {\it NuSTAR} exposure.  The spectrum
obtained by {\it NuSTAR} shows strong signatures of X-ray reflection
from the inner disk, and constrain the black hole spin parameter to be
$a\geq 0.84$.  This spin is broadly consistent with an initial report
from Gendreau et al.\ (2017), who made preliminary fits to a {\it
  NICER} observation of MAXI J1535$-$571 immediately preceding the
case analyzed in this work.  The spin of the black hole in this X-ray
binary may be the parameter of greatest interest and importance, but
the presence of an additional {\it narrow} Fe~K emission line is also
worthy of note.

The broad passband, spectral resolution, sensitivity, and throughput
of {\it NuSTAR} have made it an excellent tool for measuring black
hole spin (see, e.g., Miller et al.\ 2013, Tomsick et al.\ 2014, El
Batal et al.\ 2016).  {\it NICER} (Gendreau et al.\ 2012) has a
narrower passband, but its spectral resolution is roughly two times
sharper ($\Delta E \simeq 137$~eV at 6 keV), and it is suited to
monitoring observations in a manner that is reminiscent of {\it RXTE}.
In the near future, {\it NICER} may contribute significantly to
efforts aimed at measuring black hole spin and revealing the geometry
of the innermost accretion flow through its monitoring capability, not
merely through single observations.

\section{Observations and Data Reduction}
We consider {\it NICER} observation 1050360106, which started on 2017
September 13 at 23:58:25 (UT).  The data were processed using {\it
  NICER} software version 2018-02-22\_V002d.  The data were screened
to omit times of passage through the South Atlantic Anomaly (SAA) and
pointing offsets greater than 54\arcsec, and to only accept data
obtained with an Earth angle $\geq30^{\circ}$ above dark limb and
$\geq 45^{\circ}$ above bright limb.  After creating GTIs based on
these criteria, \textsc{niextract-events} was used to select events
with a PI value between 20 and 1200 (0.20--12.0~keV, the nominal NICER
passband) and using "EVENT\_FLAGS=bx1x000".

This procedure resulted in a net exposure of 5112~seconds, and a
nominal average count rate of 4203~${\rm counts}~ {\rm s}^{-1}$.  In
the 6.0--7.0~keV band, the average count rate is 187~${\rm counts}~
{\rm s}^{-1}$.  {\em NICER} is not an imaging instrument, so
backgrounds must be constructed from fields observed for that purpose.
Bult et al.\ (2018) estimated the {\em NICER} background at 1.5~${\rm
  counts}~ {\rm s}^{-1}$ in the 0.4-10.0~keV band.  Different
background regions, or combinations of regions, might result in a rate
that is a few times higher, but this is still a small fraction of the
count rate observed from MAXI J1535$-$571.  Background subtraction can
be important at much lower count rates, and especially at low energy.
Owing to the very high count rate observed from MAXI J1535$-$571 and
its high interstellar column, we have not subtracted a background.

\section{Analysis \& Results}
The spectrum was analyzed using XSPEC version 12.10.0 (Arnaud 1996),
using {\it NICER} instrument response files (version 1.02) and
standard data weighting.  The extremely bright nature of MAXI
J1535$-$571 obviated any need for rebinning prior to analysis within
XSPEC.  The fitting procedure minimized the $\chi^{2}$ goodness-of-fit
statistic.  Unless otherwise noted, the errors quoted in this work are
based on the values of a given parameter on the boundary of its
$1\sigma$ confidence interval.  More conservative errors are sometimes
quoted for continuum spectral fits, but $1\sigma$ errors are standard
for line spectroscopy since this permits direct estimates of line
significance; our analysis is a combination of continuum and line
spectroscopy.

Calibration residuals remain in the Si band (conservatively, 1.7--2.1
keV) owing to sharp changes in the instrumental sensitivity
curve.  Additional strong residuals remain that are
likely tied to Au edges from the mirror coatings at and around 2.2-–2.3
keV. These issues are typical for X-ray missions and Si-based
detectors, especially in the early phase of any mission, but are
particularly noticeable here due to \textit{NICER}’s high soft X-ray
throughput.  We set a lower energy bound of 2.3 keV
for our spectral analysis. Though the spectrum might nominally be
extended up to 12 keV, the upper range of any detector band is
particularly difficult to calibrate, and we set an upper threshold of
10.0 keV. As the mission calibration is refined, it may be practical
to model the data down to 0.5 keV and up to 12 keV.

We initially explored fits with absorbed single-component models,
including a multi-color disk (\texttt{diskbb}, Mitsuda et al.\ 1984)
and power-law functions.  The line-of-sight absorption was
characterized in terms of an equivalent neutral hydrogen column
density via the \texttt{tbabs} model (Wilms, Allen, \& McCray 2000).
These models are rejected by the data (power-law: $\chi^{2}/{\nu} =
17.0$, disk blackbody: $\chi^{2}/\nu = 10.8$, where $\nu$ is the
number of degrees of freedom).  A significantly improved fit is
obtained when these components are combined ($\chi^{2}/\nu = 3.1$);
the residuals are consistent with calibration uncertainties below
$E\leq 3$~keV and broad Fe~K emission, signaling reflection from the
inner disk (see Figure 1).  In these fits, there is evidence of an
edge feature at $E\simeq 9$~keV.  It is not fully consistent with the
K-shell edges of Fe XXV or Fe XXVI (8.83~keV and 9.28~keV,
respectively), nor with Au L-shell edges (9.63~keV and 9.71~keV) that
could be instrumental.  The feature is weak and does not affect the
measurements of black hole spin made below.

Strong disk reflection is expected based on the prior reports from
Gendreau et al.\ (2017) and Xu et al.\ (2018).  We therefore proceeded
to make fits with a model consisting of line-of-sight absorption in
the ISM, a disk blackbody component, a relativistically blurred
reflection model (\texttt{relxill}, which includes a cut-off
power-law; Garcia et al.\ 2014, Dauser et al.\ 2014), and a
photoelectric absorption edge.  The \texttt{relxill} model assumes a
broken power-law emissivity function (for $r\leq r_{br}$, $j\propto
r^{-q_{in}}$; for $r\geq r_{br}$, $j\propto r^{-q_{out}}$, where $j$
is the emissivity); following recent work motivated by ray-tracing
studies, we bounded the inner emissivity in the range $3 \leq q_{in}
\leq 10$, the outer emissivity in the range $0 \leq q_{out} \leq 3$,
and the break radius in the range $2~GM/c^{2} \leq r_{br} \leq
6~GM/c^{2}$ (see Wilkins \& Fabian 2012; also see Miller et al.\ 2013,
Tomsick et al.\ 2014).  Owing to the narrow fitting range, we
arbitarily fixed the power-law cut-off energy at its default value of
$E_{cut} = 300$~keV.

This model yielded a vastly improved description of the data
($\chi^{2}/\nu = 989.6/752 = 1.32$).  This fit is not formally
acceptable, largely owing to the small calibration uncertainties that
remain in the 2.3--3~keV range; even 1--2\% residuals become
significant at the count rates observed from MAXI J1535$-$571.
However, there is also evidence of a (comparatively) narrow Fe~K
emission line in the 6.6--6.7~keV range that is not modeled by
reflection from close to the ISCO (see Figures 2 and 3).  After
rebinning heavily, the line appears to be asymmetric
(if it is a single line), as per a (less-) relativistic line emerging
from a range of radii outside of the ISCO.  We therefore made fits
including the stand-alone \texttt{relline} model (the basis of the
blurring function within \texttt{relxill}).  We also explored fits with a
simple Gaussian function, and combinations of three Gaussians with
linked velocity widths, and centroids corresponding to the He-like Fe
XXV triplet.  This might correspond to lines produced in a disk wind
viewed at an intermediate inclination (see, e.g., Miller et al.\ 2015,
2016).

The best model is the one that fits the (comparatively) narrow Fe~K
emission line with a \texttt{relline} function ($\chi^{2}/\nu =
944.3/747 = 1.264$); the parameters of this model are listed in Table
1 (also see Figure 2, and Figure 3).  Fits with a single Gaussian are
significantly worse ($\chi^{2}/\nu = 969.0/749 = 1.322$, excluded at
the $4.0\sigma$ level via an F-test).  The triplet of Gaussians with
linked widths and line centroids corresponding to He-like Fe XXV is
not rigidly excluded ($\chi^{2}/\nu = 951.2/748 = 1.326$, worse than
the best-fit model at the $2.3\sigma$ level of confidence).

In our best-fit model with \texttt{relxill}, the spin of the black hole is
tightly constrained: $a = 0.994(2)$.  This is broadly consistent with
the results reported by Gendreau et al.\ (2017), and results
from an independent spectrum obtained with {\it NuSTAR} (Xu et
al.\ 2018).  The inner edge of the accretion disk is measured to sit
at a radius consistent with the ISCO: $r_{in} = 1.08(8)~r_{ISCO}$.
Other parameters critical to the reflection spectrum are also well
determined: the inner emissivity index is measured to be $q_{in} =
7.8(4)$, and the outer emissivity index is measured to be $q_{out} =
1.40^{+0.08}_{-0.19}$.  This profile is broadly consistent with
theoretical predictions of a very compact corona at a low scale height
above the black hole (e.g., Wilkins \& Fabian 2012; also see Miller et
al.\ 2013).  The measured reflection fraction of $f_{refl} =
2.65^{+0.05}_{-0.18}$ indicates that the observed spectrum is
reflection--dominated; this is also consistent with a compact corona
close to the black hole.  Were a coronal source more distant from the
black hole, light bending would be less severe, and more direct
emission would be observed.

Although the results of fits made using the most basic
\texttt{relxill} model strongly suggest a compact corona, the inferred
corona may differ from a true "lamp-post" geometry ("lamp-post"
versions of \texttt{relxill} produce significantly worse fits:
$\chi^{2}/\nu = 1062.0/749$, excluded at more than $8\sigma$ via an
F-test).  This may be echoed by the "break" radius of the emissivity
function, $r_{br} = 5.8(2) GM/c^{2}$; the radius is somewhat larger
than the value measured in other sources, and may indicate a corona
that is more radially extended.  The best-fit model also indicates a
fairly high disk temperature of $kT = 2.16(4)$~keV, and a very small
emitting area, though both are broadly consistent with values measured
in "intermediate" and "very high" states in some sources (e.g.,
Sobczak et al.\ 1999).  The power-law index, $\Gamma = 2.303(5)$, also
indicates that the source was likely observed in such a state.  Of
course, the measured values are the color temperature and area, and
standard correction factors would imply a lower temperature and larger
emitting area (e.g., Sobczak et al.\ 1999, Park et al.\ 2004).

The results of fits with \texttt{relline} are of particular interest.
The inner radius of the line production region is measured to be
$r_{in} = 144^{+140}_{-60}~GM/c^{2}$.  At such radii, the emissivity
is expected to trend toward the Euclidian value of $q=3$ (Wilkins \&
Fabian 2012); this parameter is not tightly constrained, but is
measured to be consistent with this: $q = 2.65^{1.35}_{-0.65}$.  The
inclination of the narrow line is measured to be $\theta =
37^{+22}_{-13}$ degrees; nominally, this differs from the inclination
of the inner disk measured with \texttt{relxill}, $\theta = 67.4(8)$
degrees.  This is potentially interesting as differing inclinations
may signal a warp, but the 90\% confidence
errors on the inclination include the value from the innermost disk.

The "narrow" Fe~K emission line is {\it weak}, with a measured
equivalent width of $W = 10(1)$~eV.  Yet, in this {\it NICER}
spectrum, the line is required at the $7.5\sigma$ level of confidence,
as measured by dividing the line normalization by its $1\sigma$
minus-side error.  Replacement of the \texttt{relline} model with an
unblurred reflection model, \texttt{xillver} (the core of the
\texttt{relxill}), gave a significantly worse fit ($\chi^{2}/\nu =
1116.6/747.0 = 1.495$, excluded at more than the $8\sigma$ level of
confidence).  This establishes that the line broadening is dominated
by dynamics, not scattering.

Modeling with two full relativistic reflection components can
potentially account for scattering and give an improved estimate of
the radius at which the secondary reflection occurs.  We therefore
explored fits with {\it two} full relativistic reflection models,
including the basic \texttt{relxill} and versions of that model suited
to a Comptonization continuum (\texttt{relxillCp}) and a high-density
disk (\texttt{relxilld}).  The electron temperature of the corona is
not well constrained via the Comptonization model.  The
density-sensitive models prefer a number density of $n\simeq
10^{18}~{\rm cm}^{-3}$ in the inner disk, slightly lower than recently
reported in the neutron star binary Serpens X-1 (Ludlam et al.\ 2018).
In all cases, the inner disk radius remains consistent with the ISCO
and the spin remains very high, $a\geq0.96$.  These models are
generally worse by $\Delta \chi^{2} \simeq 20-50$ than the model in
Table 1, but they all return inner radii for the second reflector in
the range of $r = 100-900~GM/c^{2}$ within 90\% confidence.

This effort revealed that there are numerous ways by which two full
relativistic reflection models might be linked or separated; the
physical implications of these choices are complex and nuanced.  A
full treatment of these matters is beyond the scope of this paper, but
the issues are important, especially at the sensitivity achieved with
{\it NICER}.  A detailed treatment is deferred to a later paper.

\section{Discussion}
We have analyzed an early {\it NICER} observation of the bright black
hole candidate, MAXI J1535$-$571.  The source was likely observed in
an "intermediate" or "very high" state, based on the observed
parameters of the continuum spectrum.  Very strong reflection from an
accretion disk extending to the ISCO is observed, and the best-fit
model measures a very high spin parameter: $a = 0.994(2)$.  This value
is compatible with prior reports (Gendreau et al.\ 2017, Xu et
al.\ 2018).  Relativistically-shaped disk reflection spectra --
including characteristic Fe~K emission lines -- are common in
stellar-mass black holes; {\it NICER} has enabled the simultaneous
detection and study of a narrower Fe~K line.  The shape of the line is
likely dominated by dynamics, and simple fits suggest that it may
originate as close as $r = 144^{+140}_{-60}~GM/c^{2}$; other fits
suggest that larger radii remain possible.  In this section, we
discuss some of the strengths and weaknesses of our spectral modeling,
implications for the innermost and intermediate accretion disk, and
the potential of future {\it NICER} observations and studies.

The spin of MAXI J1535$-$571 was measured using the \texttt{relxill}
model (Dauser et al.\ 2014, Garcia et al.\ 2014), which is now the
standard for disk reflection modeling and spin constraints.  The
unusual precision of the spin measurement reflects the sensitivity
that can be achieved using {\it NICER}.  However, we caution that
systematic uncertainties are likely to be much larger than the quoted
statistical errors.  Spin measurements using the accretion disk depend
on the extent to which the optically-thick disk is truncated at the
test-particle ISCO, making this the most important source of
systematic errors.  Currently, this question can only be explored
using numerical simulations of accretion disks; for now, at least,
these efforts suggest that fluid disks respect the test particle ISCO
for $L \leq 0.3~L_{Edd.}$ (e.g., Reynolds \& Fabian 2008; Shafee,
Narayan, \& McClintock 2008).  Remaining errors in the energy
calibration of {\it NICER} are about 30~eV; since the spin is
primarily derived from the shift of the red wing of the Fe K line,
this roughly translates to a 0.4\% systematic error on the spin parameter.

As the distance to MAXI J1535$-$571 and the mass of the black hole
primary are unknown, we cannot estimate the Eddington fraction at
which this {\it NICER} observation was made.  However, the fact that a
relativistic line characteristic of the innermost disk is clearly
detected signals that matter has not been ejected into the line of
sight, as per the case of the likely super-Eddington outburst of V404
Cyg (e.g., King et al.\ 2015).  Moreover, clear QPOs are
contemporaneously detected with coherence values of $Q=6$ (Gendreau et
al.\ 2017); scattering in a dense outflow would lead to different
light travel times and would de-cohere frequencies produced at smaller
radii.

Although it is weak, the relatively narrow Fe~K line detected between
6.6--6.7~keV may be particularly important.  A similar feature was
detected with {\it NuSTAR} (Xu et al.\ 2018), but it was not treated
with dynamical broadening.  Re-emission in a disk wind could
potentially explain this emission.  However, winds are typically
detected in ``high/soft'' or ``thermal-dominant'' states (see, e.g.,
Ponti et al.\ 2012), rather than the ``intermediate'' or ``very high''
state sampled by this spectrum.  Moreover, there is no evidence of
wind {\it absorption} in this spectrum, though winds are likely
equatorial and MAXI J1535$-$571 appears to be viewed at a high
inclination.

Our analysis favors an emission line shaped primarily by Doppler
shifts (and, weak gravitational redshifts), nominally at an
intermediate radius of $r_{in} = 144^{+140}_{-60}~GM/c^{2}$.
Preliminary fits with full reflection models allow for radii that are
larger by a factor of a few.  The modeling results are likely best
explained in terms of a warp that locally alters the profile of the
accetion disk, causing more incident radiation to be reflected.  It is
not clear if simple radiation-driven warps can persist so close to a
black hole (e.g., Pringle 1996, Maloney \& Begelman 1997).  Other
means of changing the disk profile exist; for instance,
gravitomagnetic modes can give rise to a "corrugated" disk surface
and potentially QPOs (e.g., Markovic \& Lamb 1998).

Low-frequency QPOs like those contemporaneously detected in MAXI
J1535$-$571 (Gendreau et al.\ 2017) may not be Keplerian; at minimum,
their energy dependence indicates that they are unlikely to be
produced through local dissipation in such orbits.  Reflection from
Keplerian structures at intermediate disk radii -- producing the
narrow Fe~K line in MAXI J1535$-$571 -- could connect the central
engine to distant Keplerian time scales.  Links between disk
reflection spectra and QPOs have been reported previously, and can
potentially be explained in terms of Lense-Thirring precession (see,
e.g., Miller \& Homan 2005; Miller, Homan, \& Schnittman 2006; Ingram
\& Done 2012; Ingram et al.\ 2016, 2017).

We are grateful to the anonymous referee for a helpful review of this
paper.  JMM acknowledges discussions with Phil Uttley.

\clearpage


\begin{table}[t]
\caption{Spectral Fitting Results}
\begin{footnotesize}
\begin{center}
\begin{tabular}{lll}
Component & Parameter & Value \\
\tableline
\tableline
$tbabs$ & $N_{\rm H,ISM}$ & $4.05(5)\times 10^{22}~{\rm cm}^{-2}$ \\
\tableline
$diskbb$ & kT & 2.16(4)~keV \\
--       & Norm. & $31^{+6}_{-11}$\\
\tableline
$relline$ & E & 6.72(4)~keV \\
--        & $q_{in}$ & $2.65^{1.35}_{-0.65}$ \\
--        & $q_{out}$ & $=q_{in}$ \\
--        & $r_{break}$ & $6.0^{*} GM/c^{2}$ \\
--        & $r_{in}$ & $144^{+140}_{-60}~GM/c^{2}$ \\
--        & $r_{out}$ & $990^{*} GM/c^{2}$ \\
--        & $\theta$ & $37^{+22}_{-13}$~deg. \\
--        & Norm.    & $4.5(6)\times 10^{-3}$ \\
\tableline
$relxill$ & $a = cJ/GM^{2}$ & 0.994(2) \\
--        & $q_{in}$ & 7.8(4) \\
--        & $q_{out}$ & $1.40^{+0.08}_{-0.19}$ \\
--        & $r_{break}$ & $5.8(2)~GM/c^{2}$ \\
--        & $r_{in}$ & $1.08(8)~r_{ISCO}$ \\
--        & $r_{out}$ & $990^{*} GM/c^{2}$ \\
--        & $\theta$ & $67.4(8)$~deg. \\
--        & $\Gamma$ & 2.303(5) \\
--        & log($\xi$) & 3.15(2) \\
--        & $A_{Fe}$ & 0.62(2) \\
--        & $f_{refl}$ & $2.65^{+0.05}_{-0.18}$ \\
--        & $E_{cut}$ & $300^{*}$ keV \\
--        & Norm. & 0.179(5) \\
\tableline
$edge$   & E      & 8.98(3)~keV \\
--       & $\tau$ & $5.2(5)\times 10^{-2}$ \\ 
\tableline
Total    & Flux (2.3-10~keV) & $4.5(1)\times 10^{-8}~{\rm erg}~{\rm cm}^{-2}~{\rm s}^{-1}$ \\
Total    & Unabs. flux (0.5-10~keV) & $1.47(4)\times 10^{-7}~{\rm erg}~{\rm cm}^{-2}~{\rm s}^{-1}$ \\
\tableline
\tableline
\end{tabular}
\vspace*{\baselineskip}~\\ \end{center} 
\tablecomments{The parameters of the best-fit spectral model for MAXI
  J1535$-$571.  In XSPEC parlance, the model is $tbabs\times
  (diskbb+relline+relxill)\times edge$.  The model returns
  $\chi^{2}/\nu = 944.3/747$, where $\nu$ is the number of degrees of
  freedom.  Quoted errors reflect the value of a given parameter on
  its $1\sigma$ confidence limit.  Symmetric errors in the last digit
  are quoted using parentheses.  Frozen parameters are marked with an
  asterisk.  For reference, a flux of $F\simeq 2\times 10^{-8}~{\rm
    erg}~{\rm cm}^{-2}~{\rm s}^{-1}$ is equivalent to 1~Crab in the
  2--10~keV band.  Please see the text for additional details.}
\vspace{-1.0\baselineskip}
\end{footnotesize}
\end{table}
\medskip


\clearpage


\begin{figure}
\hspace{0.75in}
\includegraphics[scale=0.5,angle=-90]{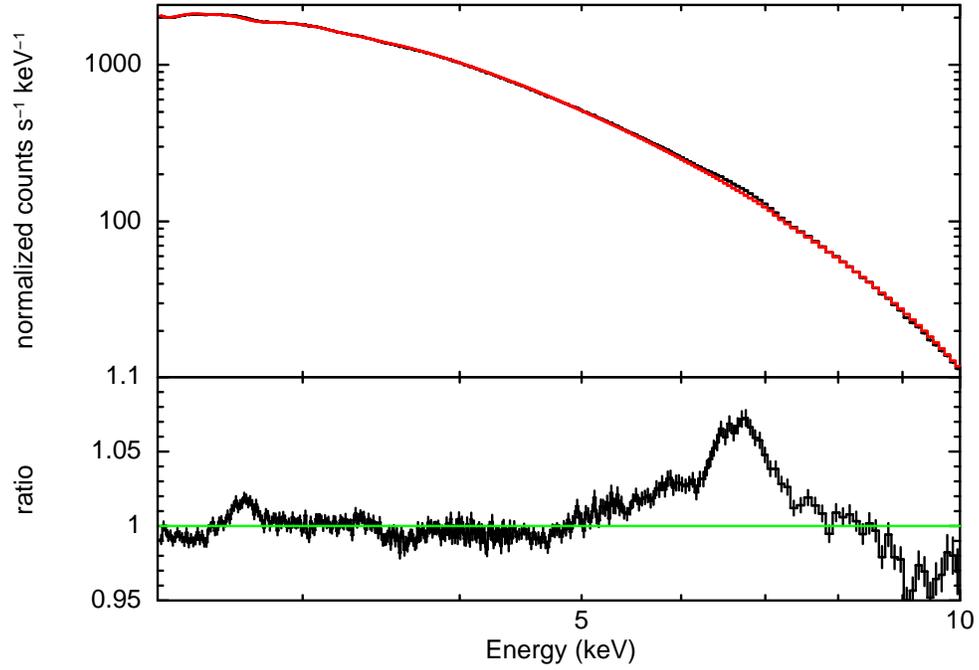}
  \figcaption[t]{\footnotesize The spectrum of MAXI J1535$-$571, fit
    with a simple disk blackbody plus power-law model.  The data are
    binned for visual clarity only, and the 4--7~keV was ignored in
    constructing the continuum fit.  The data/model ration clearly
    reveals features consistent with strong, relativistically-skewed
    reflection from the inner accretion disk.  Fits with
    \texttt{relxill} (e.g., Dauser et al.\ 2014, Garcia et al.\ 2014)
    measure a very high spin parameter, $a = cJ/GM^{2} = 0.994(2)$
    ($1\sigma$ statistical errors only).  Please see the text and
    Table 1 for additional details.}
\end{figure}
\medskip

\clearpage

\begin{figure}
\hspace{0.75in}
\includegraphics[scale=0.5,angle=-90]{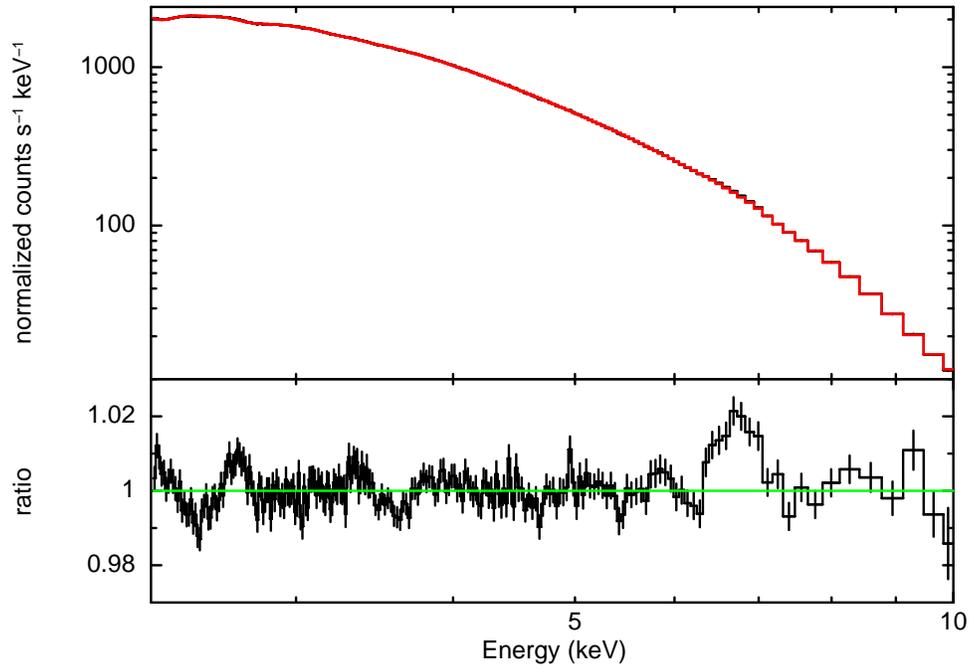}
  \figcaption[t]{\footnotesize The best-fit model for MAXI J1535$-$571
    is shown. The data have been rebinned for visual clarity only.
    The model includes blurred reflection from the inner disk, and a
    relatively narrow emission line that can be fit with a line
    originating at $r_{in} = 144^{+140}_{-60}~GM/c^{2}$.  The
    normalization of the narrow line has been set to zero in this plot
    to illustrate its nature.  Please see the text and Table 1 for
    additional details.}
\end{figure}
\medskip

\clearpage

\begin{figure}
\includegraphics[scale=0.35,angle=-90]{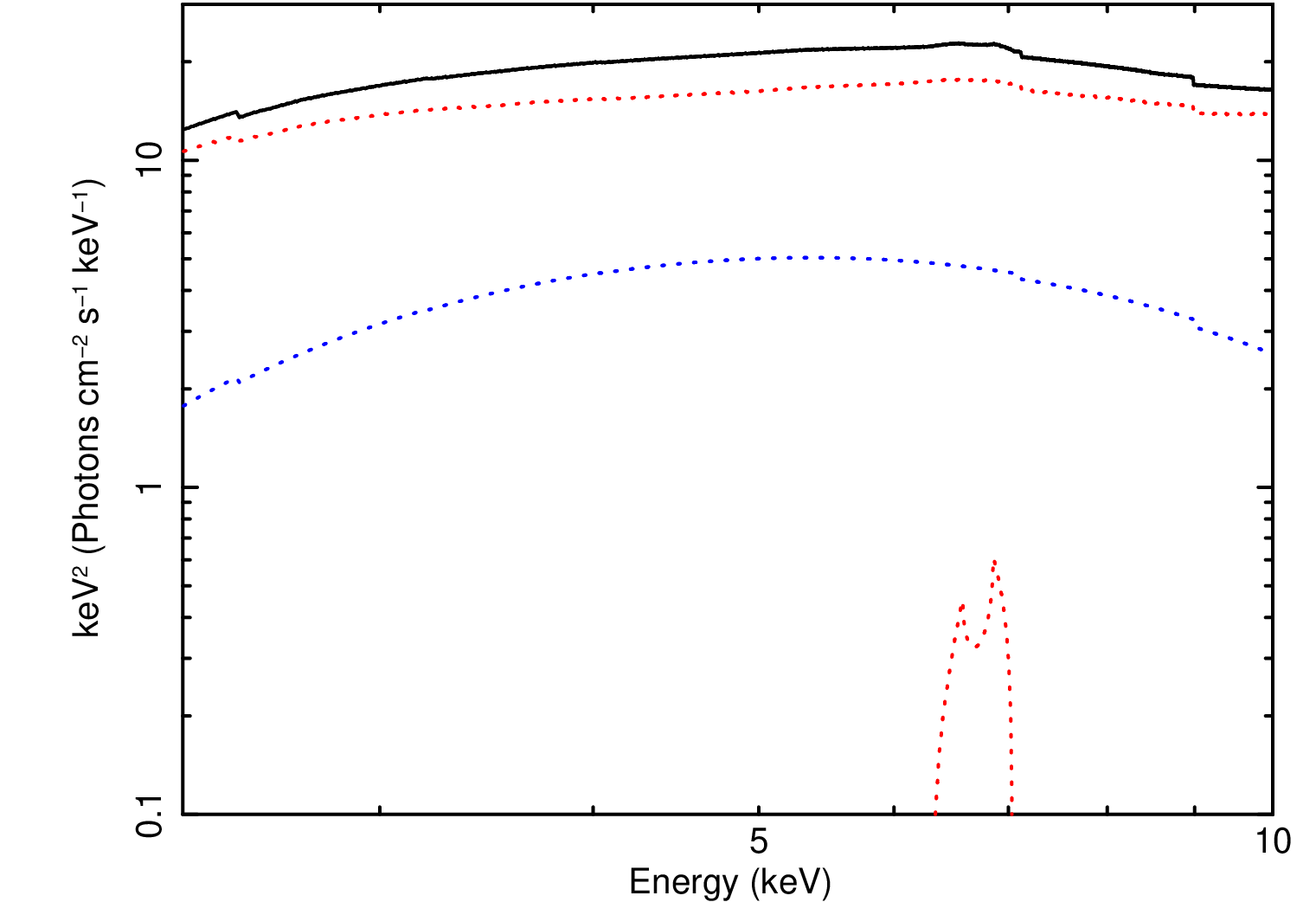}
\hspace{-0.15in}
\includegraphics[scale=0.35,angle=-90]{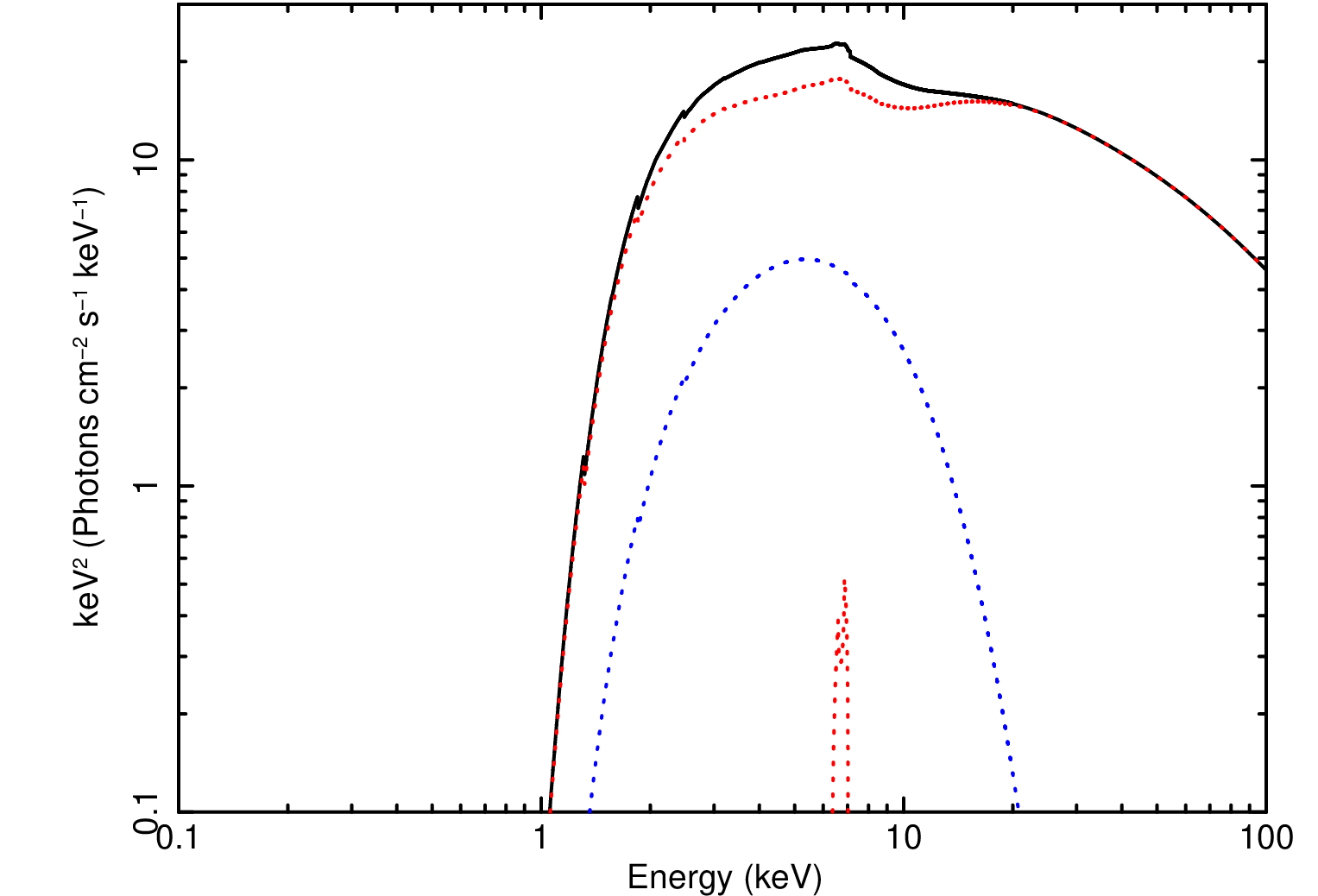}
  \figcaption[t]{\footnotesize LEFT: The best-fit model for MAXI
    J1535$-$571 is shown here without the data, in the passband of the
    fit.  The total model is shown in black; extremely blurred
    reflection from the innermost disk (including an incident
    power-law) and a separate, mildly relativistic line are shown in
    red; and the disk component is shown in blue.  RIGHT: The best-fit
    model, extrapolated over a larger range.  Please see the text and
    Table 1 for additional details.}
\end{figure}
\medskip


\end{document}